\documentclass[aps, pre, reprint, floatfix, superscriptaddress]{revtex4-1}

\usepackage{amssymb, amsmath, graphicx,verbatim, color,subfigure, natbib} 
\begin{document}
\title{Phonon and  Thermal Conducting Properties of Borocarbonitride (BCN) Nanosheets}

\author{Himanshu Chakraborty}
\affiliation{Institute for Computational Molecular Science, Center for the Computational Design of Functional Layered Materials, Temple University, Philadelphia,, PA 19122, USA.}
\altaffiliation{chakraborty.himanshu@gmail.com} 

\author{Santosh Mogurampelly}
\affiliation{Institute for Computational Molecular Science, Temple Materials Institute (TMI), 1925 North 12th St, Philadelphia, PA 19122, USA.}
\author{Vivek K. Yadav}
\affiliation{Institute for Computational Molecular Science, Department of Chemistry, Temple University, Philadelphia, PA 19122, USA.}
\author{Umesh V. Waghmare}
\affiliation{Theoretical Sciences Unit, Jawaharlal Nehru Centre for Advanced Scientific Research, P.O Jakkur, Bangalore 560064, India.}

\author{Michael L. Klein}
\altaffiliation{Mike.Klein@temple.edu} 
\affiliation{Institute for Computational Molecular Science, Center for the Computational Design of Functional Layered Materials, Temple Materials Institute (TMI), 1925 North 12th St, Philadelphia, PA 19122, USA.}

%

\date{\today}

\begin{abstract} 
Hexagonal borocarbonitrides (BCN) are a class of 2D materials, which display excellent catalytic activity for water splitting. Here, we report analysis of thermal stability, phonons and thermal conductivity of BCN monolayers over a wide range of temperatures using classical molecular dynamics simulations. Our results show that in contrast to the case of graphene and boron nitride monolayers, the out-of-plane phonons in BCN monolayers induce an asymmetry in the phonon density of states at all temperatures. Despite possessing lower thermal conducting properties compared to graphene and BN monolayers, the BCN nanosheets do not lose thermal conductivity as much as graphene and BN in the studied temperature range of 200-1000 K, and thus, the BCN nanosheets are suitable for thermal interface device applications over a wide range of temperatures. Besides their promising role in water splitting, the above results highlight the possibility of expanding the use of BCN 2D materials in thermal management applications and thermoelectrics.
\end{abstract} 

\keywords{phonons, thermal conductivity, borocarbonitrides, graphene, boronitride} 
\maketitle

\section{Introduction} 
\label{sec:introduction} 
Since the excitement generated by the extraordinary properties of graphene, \cite{novoselov-2004,geim-2007} several new 2D nanomaterials have emerged with unusual physical properties offering the possibility of novel applications. \cite{geim-2013} Among the class of hexagonal graphene (C) and boronitride (BN) 2D sheets, recently C. N. R. Rao and co-workers \cite{rao-jmca-2013} successfully synthesized hybrid hexagonal borocarbonitride (BCN) nanosheets, which contain carbon, boron and nitrogen atoms on a honeycomb lattice. 

The composition of C, B and N atoms dictates the physical properties of BCN nanosheets. For instance, the BCNs are reported to exhibit tunable band gap depending on the composition of C, B and N atoms, spanning the bandgap between zero to several eVs. \cite{rao-jmca-2013,shirodkar-BCN-2015} Therefore, the BCN materials offer superior flexibility in engineering the electrical properties of graphene-BN sheets for application in electronics. Besides, C. N. R. Rao and co-workers \cite{rao-himanshu-ees2016} demonstrated that the hydrogen evolution reaction (HER) activity is significantly enhanced with BCN nanosheets containing 20\% BN and 80\% C atomic composition. Since they are easy to synthesize, BCN nanosheets have huge potential in replacing the more expensive platinum-based conventional catalysts used in water splitting. As a result, the borocarbonitrides have generated interest in the field of materials science for their promising role in the production of clean energy. Similarly, there have been reports on the synthesis of hybrid graphene/h-BN structures, and their use in energy storage devices, field effect transistors and gas storage devices.\cite{ajayan2011,rao-jmca-2013,pati-nanoscale2014,rao-acsami-2017}  

With a goal to widen the spectrum of applications for graphene/h-BN hybrids, their thermal and related properties have been investigated theoretically. \cite{tahir2012,liu-BCN-2017,rao-BCN-2009,jindal-2016}  For example, Tahir et al. \cite{tahir2012} studied the thermal properties of the interfaces in hybrid graphene/h-BN superlattices and `dots'. The investigated hybrid materials contained smaller h-BN patches inserted as `dots' within the large graphene sheets. They found that the thermal conductivity of these hybrid materials depends sensitively on the shape and distribution of such h-BN `dots' within the graphene nanostructure.
In a different context, Liu et al \cite{liu-BCN-2017} reported thermal conductivity of hybrid graphene and h-BN nanosheets that are characterized by interfaces. Although these hybrid materials considered in the above reports \cite{tahir2012,liu-BCN-2017} share some similarities with the BCN synthesized by C. N. R. Rao et al., the phonon and thermal conducting properties of homogeneously ordered BCNs have not yet been explored. 
There also exist numerous reports studying the phonon and/or electronic properties of hybrid nanomaterials containing either B or N doped graphene using quantum mechanical calculations. \cite{rao-BCN-2009,jindal-2016} Recently, in addition to pristine BN and graphene sheets, B and N co-doped graphene has been theoretically modeled recently to explore the electronic and phonon properties. \cite{rao-BCN-2014,pati-BCN-2014,shirodkar-BCN-2015,shirodkar-BCN-2016,pati-BCN-2016,pati-BCN-2011,peeters-BN-2013,hong-2016} However, the thermal conductivity of the BCN sheets and the effects of temperature have not been investigated.

Motivated by this background, here we investigate the thermal and phonon properties of the BCNs consisting carbon, boron and nitrogen atoms arranged homogeneously on a honeycomb lattice. In addition, we also explored the temperature dependence of the phonon properties and thermal conductivity of the BCN nanosheets, and compare these findings with those of pristine graphene and 2D h-BN.  

\begin{figure}[h]
	\centering
	\includegraphics[height=95mm]{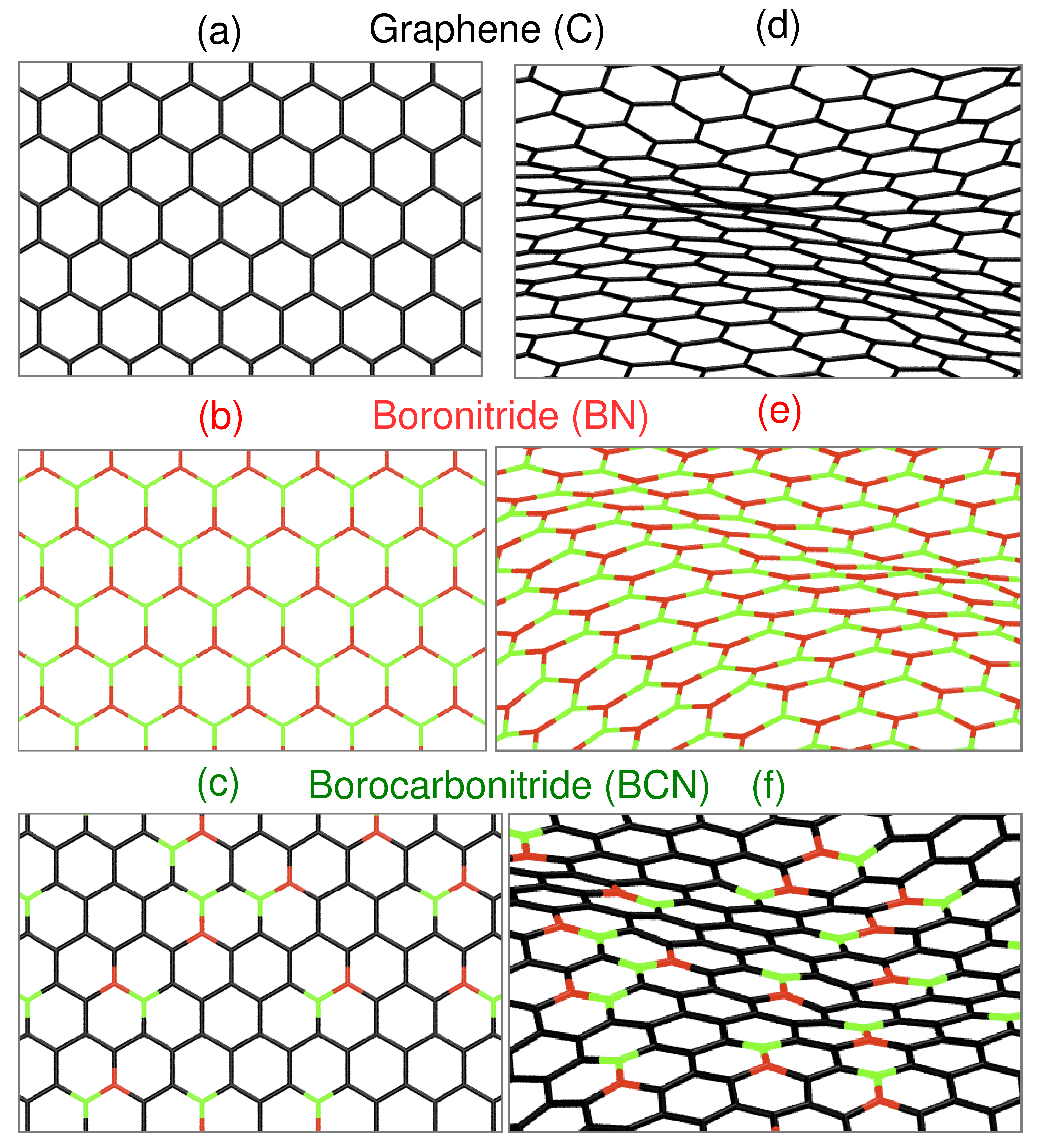}
	\caption{Schematic of various 2D monolayers investigated in this work showing the atomic composition; (a-c) represents the initial configurations and (d-f) displays the rippled structures of C, BN and BCN monolayers at room temperature. The composition of BCN consists of 80\% of C, 10\% of B and 10\% of N atoms where in the B-N bonds are distributed homogeneously. Color legend: black, red and green colors represent carbon, boron and nitrogen atoms, respectively.}
	\label{fig:fig1}
\end{figure}
In this work we used atomistic classical MD simulations to study the thermal properties of hexagonal BCN monolayers with the composition corresponding to the highest HER activity.\cite{rao-himanshu-ees2016} Specifically, we examined BCN monolayers containing 10\% B, 10\% N, and 80\% C atoms. Representative initial configurations and temperature-induced rippled structures of the graphene, BN and BCN monolayers at 300 K are shown in Fig. \ref{fig:fig1}.

\section{Simulation Details}
\label{sec:simulations}
We used optimized Tersoff force field parameters \cite{broido2010,tahir2012,koukaras-srep-2015} to describe the inter-atomic interactions relevant to BCN monolayers. We used LAMMPS MD simulation package \cite{lammps} with periodic boundary conditions considered for different system sizes between supercells 10$\times$10$\times$1 and 100$\times$100$\times$1. (see Supplementary Information, SI). To avoid interactions between periodic images of the layers, we introduced a large vacuum of 200 \AA~ along the z axis.

The initial structures were subjected to successive steepest-descent and conjugate gradient minimizations with a tolerance of $10^{-8}$ for energy and force. The systems were then heated to a desired temperature over 500 ps (250000 steps) using the Langevin thermostat in a NVT ensemble, with an integration time step of 2 fs. The lattice parameters of the triclinic simulation box were then allowed to relax for 4 ns in the isothermal-isobaric ensemble (NPT) at zero pressure with a temperature and pressure coupling constants of 0.1 and 1.0 ps, respectively.\cite{martyna-1994} 
Phonon density of states, $D(\omega)$, was calculated as a Fourier transform of the velocity autocorrelation function as:
\begin{align}
D(\omega)=\frac{1}{3Nk_B T}\int_{0}^{\infty} \frac{\langle \mathbf{v}(0)\cdot \mathbf{v}(t)\rangle}{\langle \mathbf{v}(0)\cdot \mathbf{v}(0)\rangle} e^{i\omega t} dt,
\end{align}
where, $\langle \mathbf{v}(0)\cdot \mathbf{v}(t)\rangle$ defines the velocity autocorrelation function (VACF), $\omega$ is the frequency, $N$ is the number of atoms, $k_B$ is the Boltzmann constant and $T$ is the absolute temperature. The angular bracket, $\langle$ $\cdots$ $\rangle$ indicates an ensemble average, obtained from a 50 ps long NVE trajectory generated with a finer timestep of 0.05 fs, saving velocities with a frequency of 2 fs in the computation of VACF. The finer integration timestep was used for the analysis of different time autocorrelation functions with higher resolution.  For reliable statistics, at least 20 independent NVE trajectories were generated with different initial velocities following the Maxwell-Boltzmann distribution. 

\section{Results and Discussions}
\label{sec:results} 
\subsubsection{Phonon density of states $D(\omega)$}
\begin{figure}[h]
	\centering
	\subfigure{\includegraphics[height=61mm]{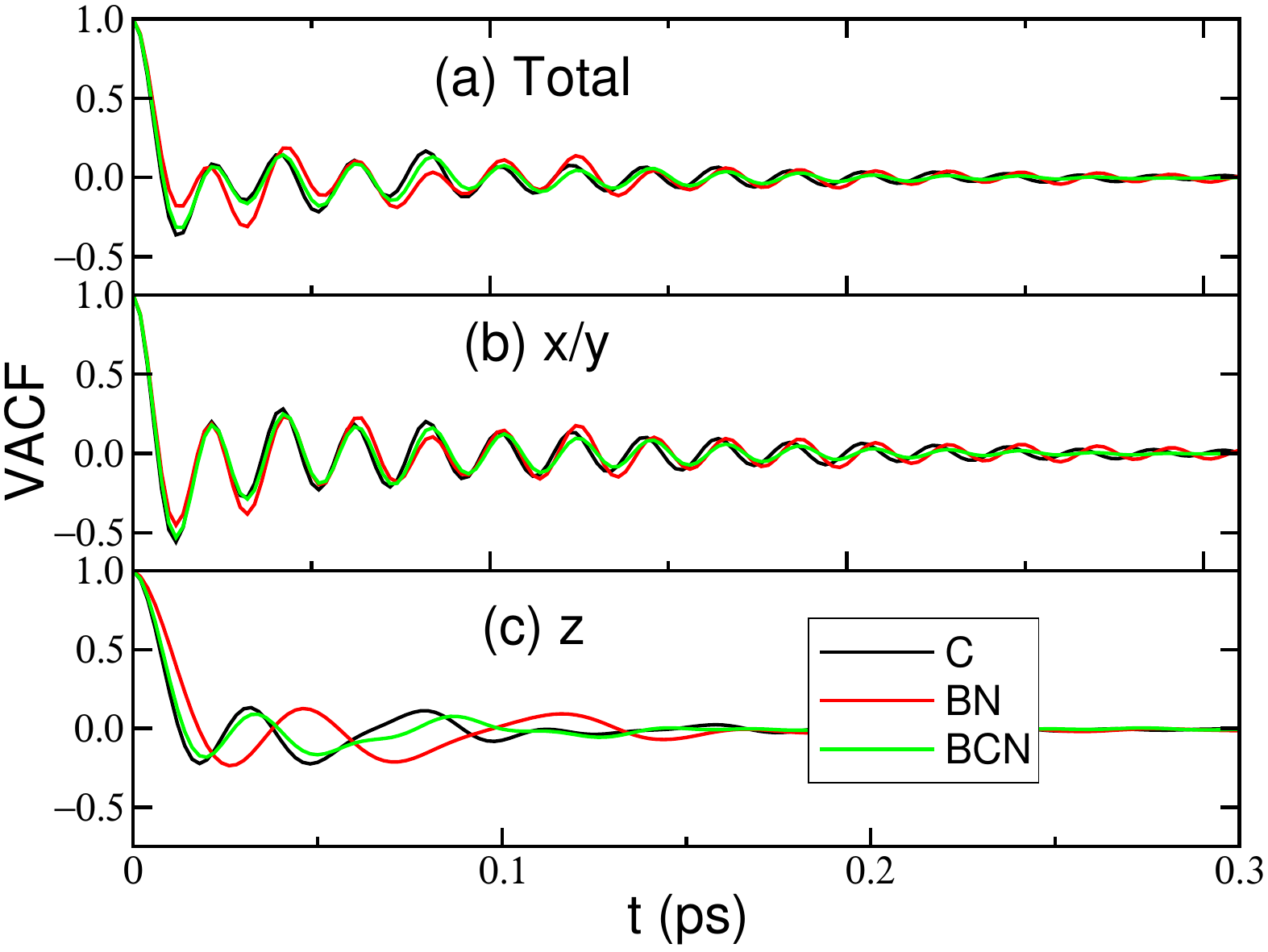}}
	\subfigure{\includegraphics[height=61mm]{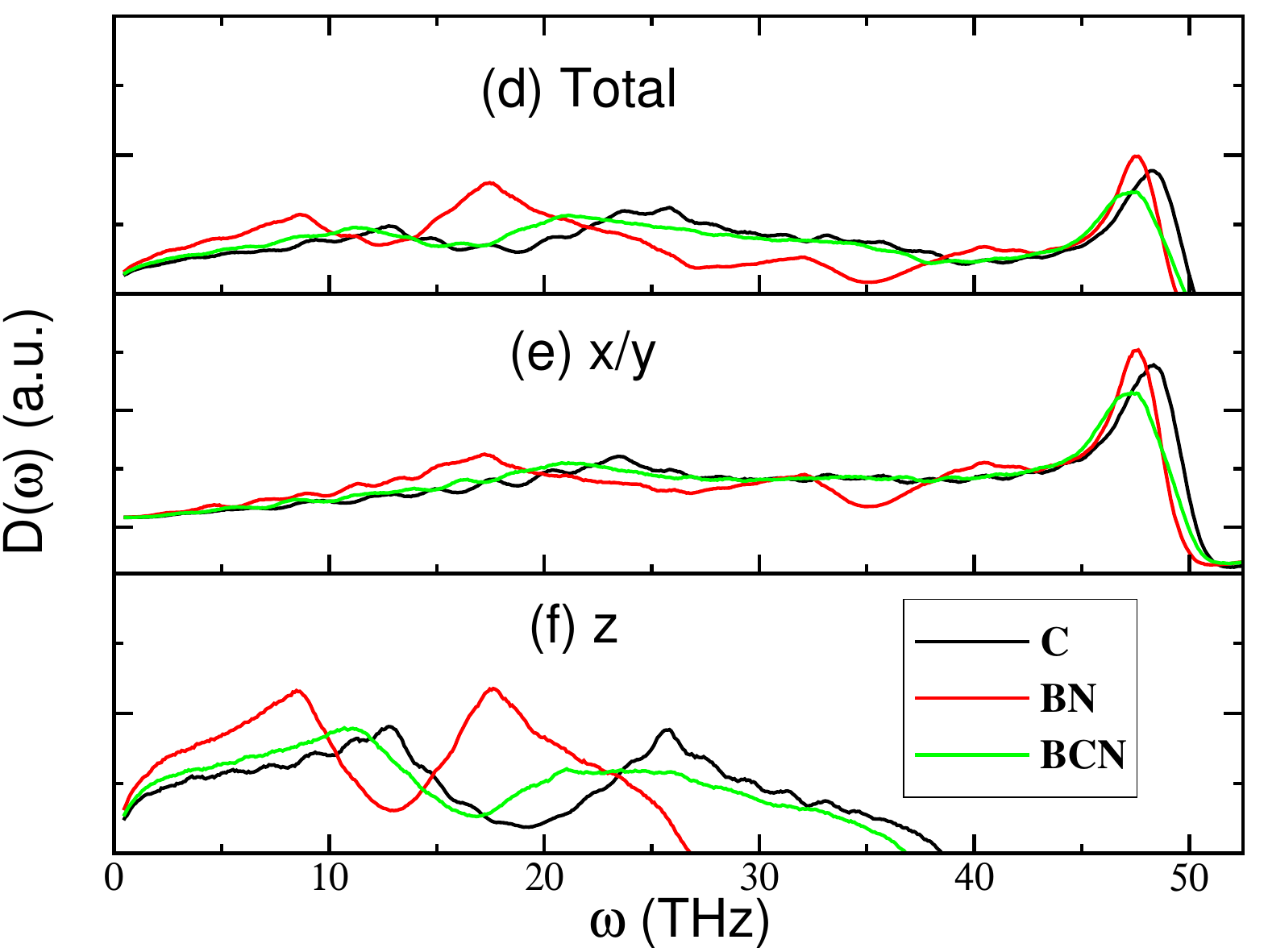}}
	\caption{(a-c) Comparison of the velocity autocorrelation function (VACF) and the (d-f) phonon density of states $D(\omega)$ for graphene, BN and BCN systems.}
	\label{fig:fig2}
\end{figure}

In the velocity autocorrelation function (VACF) of the graphene, BN and BCN monolayers at 300 K, (see Fig \ref{fig:fig2}(a)), we observe that the three nanosheets exhibit qualitatively similar behavior, with an envelope relaxation time of less than 0.11 ps. Explicitly, the location of minima and maxima of the VACF are found to be similar for these monolayers. First minima of the VACF are observed at 12 fs indicating the corresponding back scattering time scale. However, the peak value varies for different nanosheets, with the peak of BCN lying between its counterpart monolayers. As reported earlier, \cite{hong-2016} graphene and BN display distinct in-plane and out-of-plane lattice vibrations. To understand the directionality of lattice vibrations in different monolayers, we decomposed the VACF into x/y and z contributory directions and the results are presented in the Fig. \ref{fig:fig2}(b-c). Consistent with the literature \cite{hong-2016}, we find that the in-plane and out-of-plane lattice vibrations of graphene/BN (as revealed by the x/y and z components of the VACF, respectively) are remarkably different. Interestingly, the BCN nanosheets exhibit qualitatively similar behavior. 

A comparison of phonon density of states $D(\omega)$ of BCN monolayer with graphene and BN is presented in Fig. \ref{fig:fig2}d. We also present the total and  $D(\omega)$ decomposed in x/y and z directions in the Fig. \ref{fig:fig2}(e-f) respectively. The $D(\omega)$ of BCN is found to be similar to that of graphene but differs from that of BN monolayers. Such behavior is expected to arise from the contribution of carbon atoms to $D(\omega)$, which are in largest proportion with respect to B or N atoms in the BCN monolayer. More interestingly, we observe that the most intense peak of total $D(\omega)$ for the BN system is red-shifted with respect to that of graphene, consistent with previous reports. \cite{ajayan2011,ajayan-2011-nmat} This is due to the difference in masses of B and N atoms as compared to that in the graphene. \cite{valsakumar2016} Surprisingly, the most intense peak ($\omega$ $\approx$ 48 THz) of $D(\omega)$ spectra for the BCN layer is found to occur at a lower frequency than that in graphene and BN monolayers. This indicates weaker C-B and C-N bonds and softer bond-stretching frequencies in the BCN monolayer.

Consistent with the behavior of x/y and z components of the VACF, the in-plane and out-of-plane components of the $D(\omega)$ of these monolayers differ from each other. The z-component of $D(\omega)$ of BN display a higher red shift than that in graphene and BCN monolayers. A higher population of the ZA (out-of-plane acoustic) phonons is observed in the BN layer than those in the BCN and graphene layers. Specifically, the BCN spectra (see Fig. \ref{fig:fig2}f) shows asymmetry in the out- of-plane lattice vibrations (different peak heights and widths). We speculate that this is because of the ZA phonons contributed by C-B and C-N bonds in the BCN in contrast to those in the other two monolayers. Since the ZA phonons are important for the thermal conduction in graphene, \cite{seol-science-2010,lindsay-PRB-2010} it would be interesting to study how sensitively the ZA phonons, total $D(\omega)$ and the peak of $D(\omega)$ depend on temperature in these systems.

\subsubsection{Effects of Temperature on $D(\omega)$}
\begin{figure}[h]
	\centering
	\includegraphics[height=95mm]{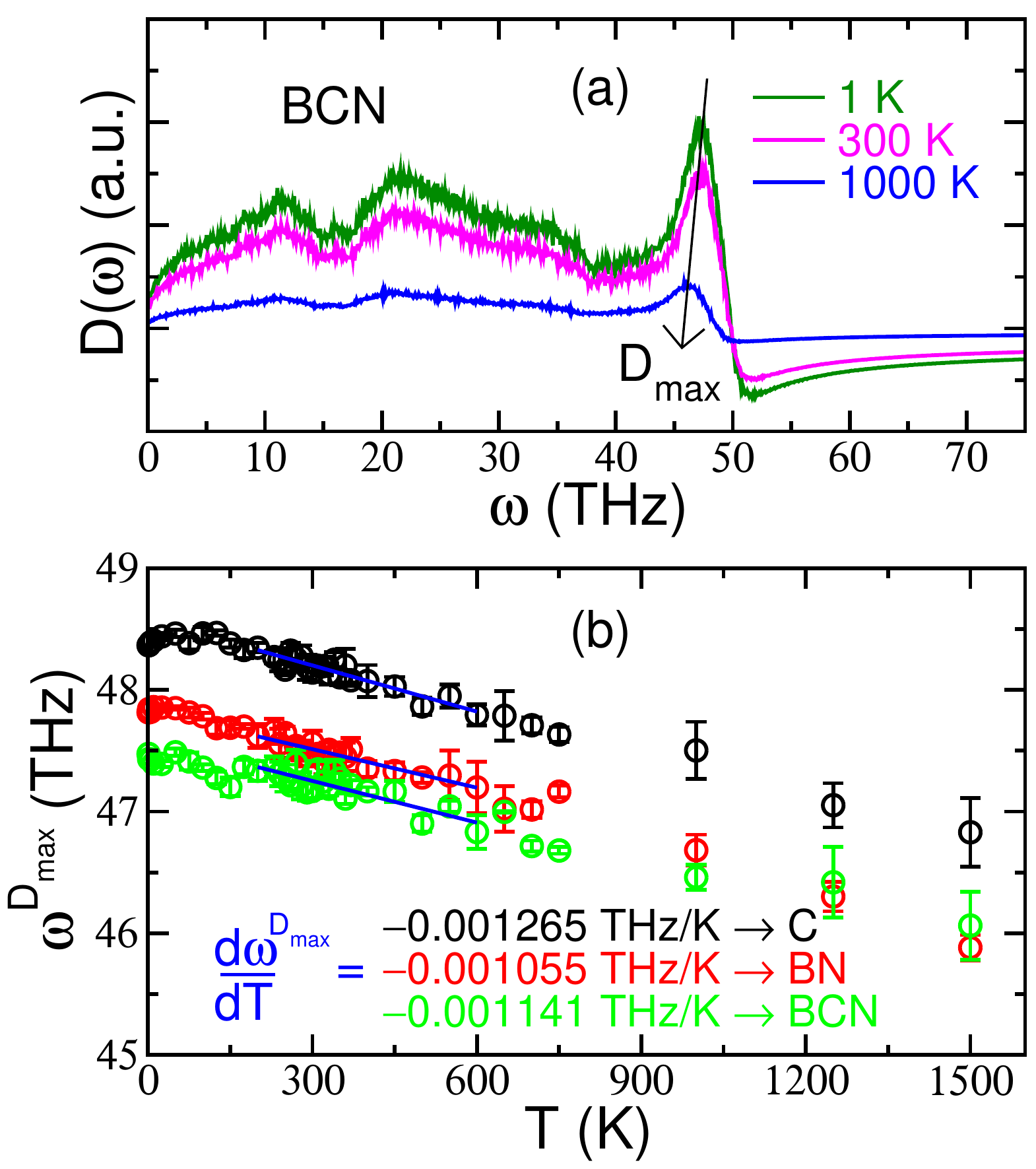}
	\caption{Effects of temperature on the (a) phonon density of states (D($\omega$)) and (b) $\omega^{\text{$D_{\text{max}}$}}$ for the graphene, BN and BCN monolayers obtained from the MD simulations. The standard deviation was calculated from at least 5 independent simulation trajectories.}
	\label{fig:fig3}
\end{figure}

In order to understand the temperature dependence of phonon properties, we carried the MD simulations for the 2D layers at various temperatures ranging from 1 K to 1500 K. Fig. \ref{fig:fig3}a presents the $D(\omega)$ of the BCN layers at three different temperatures (only the data for 1 K, 300 K and 1000 K is presented for clarity).
The frequency of the most intense peak, $\omega^{\text{$D_{\text{max}}$}}$ as a function of temperature is also presented for the different monolayers (see Fig. \ref{fig:fig3}b). As a result of thermal fluctuations, the $\omega^{\text{$D_{\text{max}}$}}$ is observed to decrease (i.e., red shifted) with temperature, typical of the softening of modes and mechanical properties of the monolayers. However, the $\omega^{\text{$D_{\text{max}}$}}$ is seen to be more sensitive to the temperature in the case of graphene monolayer with a slope, $d\omega^{\text{$D_{\text{max}}$}}$/dT of -0.00126 THz/K. Results for the temperature dependence of $D(\omega)$ and $\omega^{\text{$D_{\text{max}}$}}$ for graphene compare reasonably well with an earlier experimental report of $d\omega^{\text{$D_{\text{max}}$}}$/dT = -0.0005 THz/K. \cite{balandin-nl-2007} The quantitative discrepancy between our result and the experiments arises probably because of the Si substrate used in the reported experimental work. The BCN layer is found to have the characteristics of graphene to a larger extent with a slope of -0.00114 THz/K, while the BN sheets display the lowest variation in $\omega^{\text{$D_{\text{max}}$}}$ with temperature. Anharmonic coupling of phonon modes causes the shifts in mode frequencies, $\omega^{\text{$D_{\text{max}}$}}$  with varying temperatures for the three monolayers. Balandin and co-workers \cite{balandin-nl-2007} suggested that the red shift in the G band frequency (or $\omega^{\text{$D_{\text{max}}$}}$) in graphene single layer is mainly due to phonon-phonon coupling. Our simulations predict that the temperature coefficient $d\omega^{\text{$D_{\text{max}}$}}$/dT is the highest for graphene and lowest for BN, and the temperature dependence of $\omega^{\text{$D_{\text{max}}$}}$ for BCN is similar to that of graphene. Further, it would be interesting to understand variation of thermal conductivity with temperatures in these systems.

\subsubsection{Thermal conductivity}
In this section, we present results for the thermal conductivity and examine the influence of temperature. The thermal conductivity of various monolayers was calculated by using the Green-Kubo relation based on the fluctuation-dissipation theorem. \cite{zwanzig-1965} Specifically, the time correlation function of heat current operator was used to calculate the thermal conductivity as: \cite{McQuarrie_statmech_book}
\begin{align}
\kappa(T)=\frac{1}{Vk_B T^2}\int_{0}^{\infty} \langle \mathbf{S}(0)\cdot \mathbf{S}(t)\rangle dt,
\label{eq:kappa}
\end{align}
where $V$ is the volume of the simulation box which was computed as $3.35\times L_x\times L_y$, where 3.35 \AA~is the thickness of graphene \cite{xczeng2015} and $L_x$ and $L_y$ are the box dimensions in $x$ and $y$ directions, respectively. Since the out-of-plane contribution is least important for 2D nanosheets considered in this work, we calculate the thermal conductivity as the mean of in-plane components such that $\kappa=(\kappa_{xx}+\kappa_{yy})/2$. In the above, the heat current operator $\mathbf{S}(t)$ is given by: \cite{zwanzig-1965,lee-1991}
\begin{align}
\mathbf{S}(t)= \frac{d}{dt}\sum_i \mathbf{r}_i \tilde{E}_i,
\end{align}
where $\mathbf{r}_i$ is the position vector of $i^{\textbf{th}}$ atom and $\tilde{E}_i=E_i-\langle E_i\rangle $ is the corresponding deviation of the total energy from its average value. In our simulations, the $\mathbf{S}(t)$ was calculated using the following formula:
\begin{align}
\mathbf{S}(t)=\sum_{i}^{}\tilde{E}_i\mathbf{v}_i+\frac{1}{2}\sum_{i<j}^{}\left(\mathbf{f}_{ij}\cdot(\mathbf{v}_{i}+\mathbf{v}_{j})\right)\mathbf{r}_{ij},
\end{align}
where $\mathbf{f}_{ij}$ is the force between atoms $i$ and $j$, $\mathbf{v}_{i}$ is the velocity of $i^{\text{th}}$ atom and $\mathbf{r}_{ij}$ is the inter-particle separation vector. We note that unlike pairwise interactions, the energy due to the 3-body Tersoff potential, $V_{ijk}$ can not be uniquely assigned to any of the atoms $i$, $j$ and $k$. However, for simplicity, we compute the atomic site total energy as implemented in LAMMPS \cite{lammps} and assign in equal proportions to interacting atoms $i$, $j$ and $k$. This choice is expected to be reasonable with comparable sizes of B, C, N atoms and was inspired by the fact that temperature gradient varies on length scales larger than the interatomic distances, and the results are expected to be independent of the above choice. \cite{lee-1991} 

\begin{figure}[h]
	\centering
	\includegraphics[width=110mm]{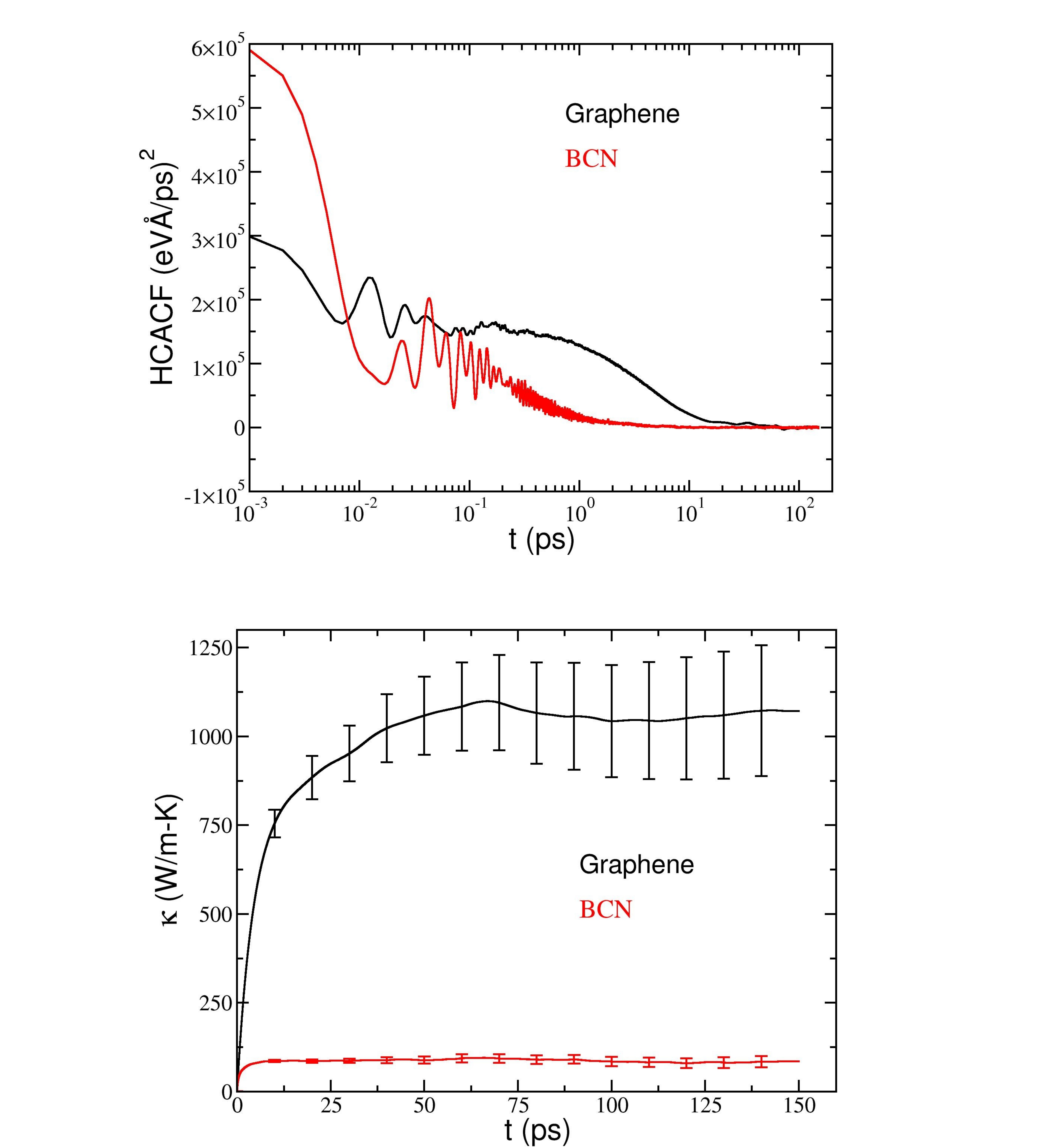}
	\caption{(a) Heat current autocorrelation function for graphene and BCN at 300 K for a system of supercell 100$\times$100$\times$1 (20000 atoms) averaged over 20 independent simulation runs of each 1 ns length and a saving frequency of 1 fs. The HCACF displays initial rapid and long time decay at different timescales. (b) Thermal conductivity calculated as numerical integration of the HCACF curves as a function of the upper cutoff for integration (see Eq. \ref{eq:kappa}) from 20 independent simulation runs.}  
	\label{fig:fig4}
\end{figure}
We note that the thermal conductivity calculations are computationally challenging in equilibrium MD simulations using Green-Kubo method because of issues such as large deviations from the average $\kappa$ (Fig. 4b), problems underlying the convergence, system size dependency etc. To overcome these problems, we have tested our computational approach by considering multiple independent simulation runs, the upper time limit appearing in the integration of Eq. \ref{eq:kappa} and the system size dependency. Specifically, we performed 20 independent simulation runs of BCN monolayers (and compared with graphene) and the average of heat current autocorrelation function (HCACF) and $\kappa$ as a function of the upper time limit for the integration are displayed in Fig. \ref{fig:fig4} (see SI for more information). It is observed that the HCACFs decay rapidly at lower times corresponding to atomic collision timescales, accompanied by slower relaxation at long timescales. \cite{tahirjcp2014} As shown in Fig. \ref{fig:fig4}(b), the thermal conductivity increases with the upper cutoff of the integration time limit and converges to a constant value after approximately 50 ps. However, we considered 150 ps as the upper limit for the HCACF integration while calculating the thermal conductivity. In addition to the convergence of $\kappa$ with the upper integration time limit, we also investigated the system size dependency on $\kappa$ and find converged values for systems equivalent to 20000 atoms or more (See Fig. S2 of SI).\cite{size-dependence-tersoff-2015}    
\begin{figure}[h]
	\centering
	\includegraphics[width=80mm]{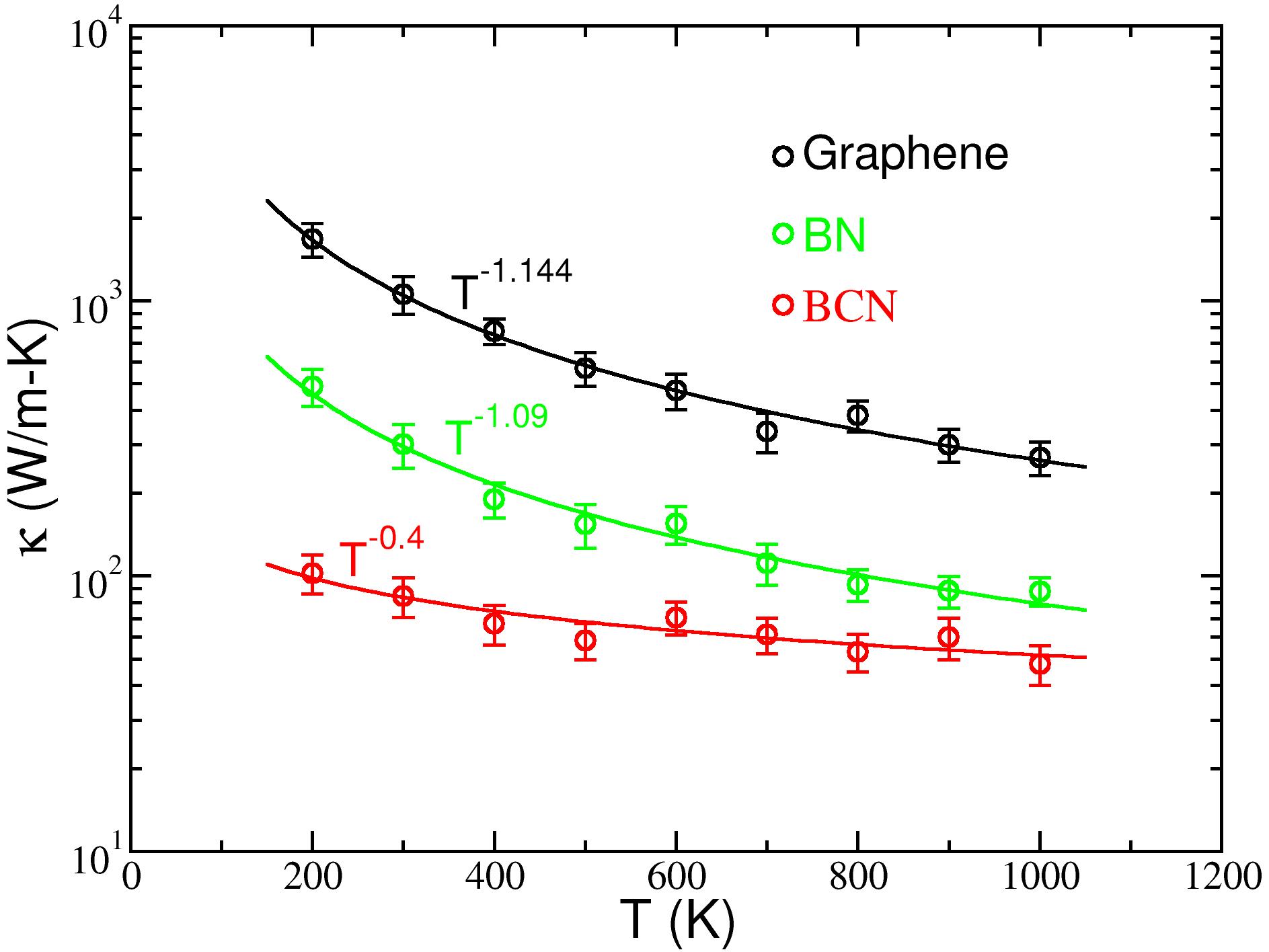}
	\caption{Thermal conductivity ($\kappa$) of BCN monolayers in comparison with graphene and BN as a function of temperature calculated from classical MD simulations. The error bar indicates the standard deviation to the mean value of $\kappa$.}  
	\label{fig:fig5}
\end{figure}

The temperature dependence of $\kappa$ for a system size of supercell 100$\times$100$\times$1 (20000 atoms, 25.1 nm $\times$ 21.7 nm) is displayed in Fig. \ref{fig:fig5}. At room temperature, we obtained a value of $\kappa$=85 $\pm$ 14 W/m-K for BCN monolayers which is lower than those of graphene ($\kappa$=1057 $\pm$ 165 W/m-K) and BN ($\kappa$=301 $\pm$ 55 W/m-K) monolayers. We note that $\kappa$ of graphene is in close agreement with Cagin and co-workers \cite{tahir2012} who employed the Tersoff parameters as in our simulations. Moreover, the $\kappa$ of BN monolayer compare well those of BN nanoribbons as reported by Khan et al. \cite{KhanBN2017}  In BCN monolayer, the phonon modes associated with C-B, C-N and B-N heteropolar bonds give rise to a rapid decay of HCACF compared to graphene (approximately at 5 ps vs 50 ps as can be seen from Fig. \ref{fig:fig4}). These differences arise from different amounts of energy carried by acoustic phonons, and, the difference in the masses of C, B and N atoms. Consistent with relative differences between timescales at which the HCACF decays to zero (5 ps vs 50 ps), thermal conductivity of BCN monolayer was found to be an order lower than that of graphene at room temperature. The presence of heteropolar bonds (C-B, C-N and B-N) increases phonon-phonon scattering and thus, lower thermal conductivity of BCN as compared to those in graphene and BN nanosheets (see Fig. \ref{fig:fig5}).
As the temperature is increased, we find a monotonic decrease in $\kappa$ for all the monolayers investigated. Interestingly, the rate of decay (with temperature) is much smaller for BCN monolayers ($\kappa(T)\sim T^{-0.4}$) compared to graphene ($\kappa(T)\sim T^{-1.14}$) and BN ($\kappa(T)\sim T^{-1.09}$) sheets which has promising consequences in thermal management applications. Specifically, the BCN monolayers can be used for thermal applications for a wide range of temperature without losing much in the thermal conducting properties, where as the graphene and BN materials performs less efficiently (highly sensitive to temperature) in similar conditions.

The results presented in this report are purely based on \textit{classical} MD simulations, which predict diverging $\kappa$ at low $T$, which is however unphysical since the phonons exhibit quantum statistical behavior at low temperatures such that $T\ll\theta$, where $\theta$ is the Debye temperature. \cite{wang-PRB-1990,tahir-JCP-2000} We note that the Debye temperature of the monolayers considered in this work is much greater than room temperature. For instance, $\theta(=1810$ K for graphene\cite{tahir-JCP-2000}) being considerably larger than the room temperature requires quantum corrections in the low temperature range ($T\ll\theta$), which will be reported in a future communication.

\section{Conclusions}
In summary, we have compared theoretical analysis of phonon density of states and thermal conductivity of BCN (with atomic composition of 80\% C, 10\% B, and 10\% N), with those of counterpart graphene and BN monolayers of the same dimensions as a function of temperature. Our work is based on the \textit{classical} molecular dynamics simulations using the 3-body Tersoff interaction parameters. The most intense peak of the phonon density of states for BCN sheets is observed to exhibit a red shift with respect to both the graphene and BN monolayers. Furthermore, the out-of-plane phonon modes cause asymmetry in the phonon density of states of BCN, in contrast to h-BN and graphene monolayers. The frequency corresponding to the most intense peak of the phonon density of states decreases with temperature. Specifically, the phonon softening frequency ($\omega^{\text{D}_{\text{max}}}$) decays linearly with a slope of $d\omega^{\text{D}_{\text{max}}}$/dT= -0.00114 THz/K for BCN in comparison with a slope of -0.00126 THz/K for graphene. We find that the thermal conductivity of BCN is about one order lower in magnitude than that of graphene and BN monolayers at room temperature due to increased phonon-phonon scattering caused by the heteropolar bonds such as C-B, C-N and B-N in the BCN nanosheet. The thermal conductivity of all the monolayers considered in this work was observed to drop rapidly as the temperature is increased (in the range 200-1000 K) with a power law $\kappa(T)\sim T^{-\lambda}$. We find that $\lambda$ is much smaller for BCN monolayers than graphene and BN monolayers indicating the suitability of BCN nanosheets over a wide range of temperatures, opening new avenues of applications in thermoelectrics and thermal interface materials.

\section{Acknowledgments}
This research includes calculations carried out on Temple University's HPC resources and thus was supported in part by the National Science Foundation through major research instrumentation grant number 1625061 and by the US Army Research Laboratory under contract number W911NF-16-2-0189. U.V.W. and M.L.K. thank HH Sheikh Saud bin Saqr al Qasimi for support via a Sheikh Saqr Research Fellowship.  U.V.W. acknowledges support from a JC Bose National Fellowship, the India-Korea Science and Technology Center and an AOARD project no. FA 2386-15-1-0002. H.C., M.L.K., and part of the computational resources were supported as part of the Center for the Computational Design of Functional Layered Materials, an Energy Frontier Research Center funded by the U.S. Department of Energy, Office of Science, Basic Energy Sciences under Award DE-SC0012575.

\bibliography{ref.bib}

\begin{thebibliography}{38}%
\makeatletter
\providecommand \@ifxundefined [1]{%
 \@ifx{#1\undefined}
}%
\providecommand \@ifnum [1]{%
 \ifnum #1\expandafter \@firstoftwo
 \else \expandafter \@secondoftwo
 \fi
}%
\providecommand \@ifx [1]{%
 \ifx #1\expandafter \@firstoftwo
 \else \expandafter \@secondoftwo
 \fi
}%
\providecommand \natexlab [1]{#1}%
\providecommand \enquote  [1]{``#1''}%
\providecommand \bibnamefont  [1]{#1}%
\providecommand \bibfnamefont [1]{#1}%
\providecommand \citenamefont [1]{#1}%
\providecommand \href@noop [0]{\@secondoftwo}%
\providecommand \href [0]{\begingroup \@sanitize@url \@href}%
\providecommand \@href[1]{\@@startlink{#1}\@@href}%
\providecommand \@@href[1]{\endgroup#1\@@endlink}%
\providecommand \@sanitize@url [0]{\catcode `\\12\catcode `\$12\catcode
  `\&12\catcode `\#12\catcode `\^12\catcode `\_12\catcode `\%12\relax}%
\providecommand \@@startlink[1]{}%
\providecommand \@@endlink[0]{}%
\providecommand \url  [0]{\begingroup\@sanitize@url \@url }%
\providecommand \@url [1]{\endgroup\@href {#1}{\urlprefix }}%
\providecommand \urlprefix  [0]{URL }%
\providecommand \Eprint [0]{\href }%
\providecommand \doibase [0]{http://dx.doi.org/}%
\providecommand \selectlanguage [0]{\@gobble}%
\providecommand \bibinfo  [0]{\@secondoftwo}%
\providecommand \bibfield  [0]{\@secondoftwo}%
\providecommand \translation [1]{[#1]}%
\providecommand \BibitemOpen [0]{}%
\providecommand \bibitemStop [0]{}%
\providecommand \bibitemNoStop [0]{.\EOS\space}%
\providecommand \EOS [0]{\spacefactor3000\relax}%
\providecommand \BibitemShut  [1]{\csname bibitem#1\endcsname}%
\let\auto@bib@innerbib\@empty
\bibitem [{\citenamefont {Novoselov}\ \emph {et~al.}(2004)\citenamefont
  {Novoselov}, \citenamefont {Geim}, \citenamefont {Morozov}, \citenamefont
  {Jiang}, \citenamefont {Zhang}, \citenamefont {Dubonos}, \citenamefont
  {Grigorieva},\ and\ \citenamefont {Firsov}}]{novoselov-2004}%
  \BibitemOpen
  \bibfield  {author} {\bibinfo {author} {\bibfnamefont {K.~S.}\ \bibnamefont
  {Novoselov}}, \bibinfo {author} {\bibfnamefont {A.~K.}\ \bibnamefont {Geim}},
  \bibinfo {author} {\bibfnamefont {S.~V.}\ \bibnamefont {Morozov}}, \bibinfo
  {author} {\bibfnamefont {D.}~\bibnamefont {Jiang}}, \bibinfo {author}
  {\bibfnamefont {Y.}~\bibnamefont {Zhang}}, \bibinfo {author} {\bibfnamefont
  {S.~V.}\ \bibnamefont {Dubonos}}, \bibinfo {author} {\bibfnamefont {I.~V.}\
  \bibnamefont {Grigorieva}}, \ and\ \bibinfo {author} {\bibfnamefont {A.~A.}\
  \bibnamefont {Firsov}},\ }\href@noop {} {\bibfield  {journal} {\bibinfo
  {journal} {Science}\ }\textbf {\bibinfo {volume} {306}},\ \bibinfo {pages}
  {666} (\bibinfo {year} {2004})}\BibitemShut {NoStop}%
\bibitem [{\citenamefont {Geim}\ and\ \citenamefont
  {Novoselov}(2007)}]{geim-2007}%
  \BibitemOpen
  \bibfield  {author} {\bibinfo {author} {\bibfnamefont {A.~K.}\ \bibnamefont
  {Geim}}\ and\ \bibinfo {author} {\bibfnamefont {K.~S.}\ \bibnamefont
  {Novoselov}},\ }\href@noop {} {\bibfield  {journal} {\bibinfo  {journal}
  {Nat. Mater.}\ }\textbf {\bibinfo {volume} {6}},\ \bibinfo {pages} {183}
  (\bibinfo {year} {2007})}\BibitemShut {NoStop}%
\bibitem [{\citenamefont {Geim}\ and\ \citenamefont
  {Grigorieva}(2013)}]{geim-2013}%
  \BibitemOpen
  \bibfield  {author} {\bibinfo {author} {\bibfnamefont {A.~K.}\ \bibnamefont
  {Geim}}\ and\ \bibinfo {author} {\bibfnamefont {I.~V.}\ \bibnamefont
  {Grigorieva}},\ }\href@noop {} {\bibfield  {journal} {\bibinfo  {journal}
  {Nature}\ }\textbf {\bibinfo {volume} {499}},\ \bibinfo {pages} {419}
  (\bibinfo {year} {2013})}\BibitemShut {NoStop}%
\bibitem [{\citenamefont {Kumar}\ \emph {et~al.}(2013)\citenamefont {Kumar},
  \citenamefont {Moses}, \citenamefont {Pramoda}, \citenamefont {Shirodkar},
  \citenamefont {Mishra}, \citenamefont {Waghmare}, \citenamefont
  {Sundaresan},\ and\ \citenamefont {Rao}}]{rao-jmca-2013}%
  \BibitemOpen
  \bibfield  {author} {\bibinfo {author} {\bibfnamefont {N.}~\bibnamefont
  {Kumar}}, \bibinfo {author} {\bibfnamefont {K.}~\bibnamefont {Moses}},
  \bibinfo {author} {\bibfnamefont {K.}~\bibnamefont {Pramoda}}, \bibinfo
  {author} {\bibfnamefont {S.~N.}\ \bibnamefont {Shirodkar}}, \bibinfo {author}
  {\bibfnamefont {A.~K.}\ \bibnamefont {Mishra}}, \bibinfo {author}
  {\bibfnamefont {U.~V.}\ \bibnamefont {Waghmare}}, \bibinfo {author}
  {\bibfnamefont {A.}~\bibnamefont {Sundaresan}}, \ and\ \bibinfo {author}
  {\bibfnamefont {C.~N.~R.}\ \bibnamefont {Rao}},\ }\href@noop {} {\bibfield
  {journal} {\bibinfo  {journal} {J. Mater. Chem. A}\ }\textbf {\bibinfo
  {volume} {1}},\ \bibinfo {pages} {5806} (\bibinfo {year} {2013})}\BibitemShut
  {NoStop}%
\bibitem [{\citenamefont {Shirodkar}\ \emph {et~al.}(2015)\citenamefont
  {Shirodkar}, \citenamefont {Waghmare}, \citenamefont {Fisher},\ and\
  \citenamefont {Grau-Crespo}}]{shirodkar-BCN-2015}%
  \BibitemOpen
  \bibfield  {author} {\bibinfo {author} {\bibfnamefont {S.~N.}\ \bibnamefont
  {Shirodkar}}, \bibinfo {author} {\bibfnamefont {U.~V.}\ \bibnamefont
  {Waghmare}}, \bibinfo {author} {\bibfnamefont {T.~S.}\ \bibnamefont
  {Fisher}}, \ and\ \bibinfo {author} {\bibfnamefont {R.}~\bibnamefont
  {Grau-Crespo}},\ }\href@noop {} {\bibfield  {journal} {\bibinfo  {journal}
  {Phys. Chem. Chem. Phys.}\ }\textbf {\bibinfo {volume} {17}},\ \bibinfo
  {pages} {13547} (\bibinfo {year} {2015})}\BibitemShut {NoStop}%
\bibitem [{\citenamefont {Chhetri}\ \emph {et~al.}(2016)\citenamefont
  {Chhetri}, \citenamefont {Maitra}, \citenamefont {Chakraborty}, \citenamefont
  {Waghmare},\ and\ \citenamefont {Rao}}]{rao-himanshu-ees2016}%
  \BibitemOpen
  \bibfield  {author} {\bibinfo {author} {\bibfnamefont {M.}~\bibnamefont
  {Chhetri}}, \bibinfo {author} {\bibfnamefont {S.}~\bibnamefont {Maitra}},
  \bibinfo {author} {\bibfnamefont {H.}~\bibnamefont {Chakraborty}}, \bibinfo
  {author} {\bibfnamefont {U.~V.}\ \bibnamefont {Waghmare}}, \ and\ \bibinfo
  {author} {\bibfnamefont {C.~N.~R.}\ \bibnamefont {Rao}},\ }\href@noop {}
  {\bibfield  {journal} {\bibinfo  {journal} {Energy Environ. Sci.}\ }\textbf
  {\bibinfo {volume} {9}},\ \bibinfo {pages} {95} (\bibinfo {year}
  {2016})}\BibitemShut {NoStop}%
\bibitem [{\citenamefont {Liu}\ \emph {et~al.}(2011)\citenamefont {Liu},
  \citenamefont {Song}, \citenamefont {Zhao}, \citenamefont {Huang},
  \citenamefont {Ma}, \citenamefont {Zhang}, \citenamefont {Lou},\ and\
  \citenamefont {M.~Ajayan}}]{ajayan2011}%
  \BibitemOpen
  \bibfield  {author} {\bibinfo {author} {\bibfnamefont {Z.}~\bibnamefont
  {Liu}}, \bibinfo {author} {\bibfnamefont {L.}~\bibnamefont {Song}}, \bibinfo
  {author} {\bibfnamefont {S.}~\bibnamefont {Zhao}}, \bibinfo {author}
  {\bibfnamefont {J.}~\bibnamefont {Huang}}, \bibinfo {author} {\bibfnamefont
  {L.}~\bibnamefont {Ma}}, \bibinfo {author} {\bibfnamefont {J.}~\bibnamefont
  {Zhang}}, \bibinfo {author} {\bibfnamefont {J.}~\bibnamefont {Lou}}, \ and\
  \bibinfo {author} {\bibfnamefont {P.}~\bibnamefont {M.~Ajayan}},\ }\href@noop
  {} {\bibfield  {journal} {\bibinfo  {journal} {Nano Lett.}\ }\textbf
  {\bibinfo {volume} {11}},\ \bibinfo {pages} {2032} (\bibinfo {year}
  {2011})}\BibitemShut {NoStop}%
\bibitem [{\citenamefont {Banerjee}\ and\ \citenamefont
  {Pati}(2014{\natexlab{a}})}]{pati-nanoscale2014}%
  \BibitemOpen
  \bibfield  {author} {\bibinfo {author} {\bibfnamefont {S.}~\bibnamefont
  {Banerjee}}\ and\ \bibinfo {author} {\bibfnamefont {S.~K.}\ \bibnamefont
  {Pati}},\ }\href@noop {} {\bibfield  {journal} {\bibinfo  {journal}
  {Nanoscale}\ }\textbf {\bibinfo {volume} {6}},\ \bibinfo {pages} {13430}
  (\bibinfo {year} {2014}{\natexlab{a}})}\BibitemShut {NoStop}%
\bibitem [{\citenamefont {Rao}\ and\ \citenamefont
  {Gopalakrishnan}(2017)}]{rao-acsami-2017}%
  \BibitemOpen
  \bibfield  {author} {\bibinfo {author} {\bibfnamefont {C.~N.~R.}\
  \bibnamefont {Rao}}\ and\ \bibinfo {author} {\bibfnamefont {K.}~\bibnamefont
  {Gopalakrishnan}},\ }\href@noop {} {\bibfield  {journal} {\bibinfo  {journal}
  {ACS Appl. Mater. Interface}\ }\textbf {\bibinfo {volume} {9}},\ \bibinfo
  {pages} {19478} (\bibinfo {year} {2017})},\ \bibinfo {note} {pMID:
  27797466}\BibitemShut {NoStop}%
\bibitem [{\citenamefont {Kinaci}\ \emph {et~al.}(2012)\citenamefont {Kinaci},
  \citenamefont {Haskins}, \citenamefont {Sevik},\ and\ \citenamefont
  {Cagin}}]{tahir2012}%
  \BibitemOpen
  \bibfield  {author} {\bibinfo {author} {\bibfnamefont {A.}~\bibnamefont
  {Kinaci}}, \bibinfo {author} {\bibfnamefont {J.~B.}\ \bibnamefont {Haskins}},
  \bibinfo {author} {\bibfnamefont {C.}~\bibnamefont {Sevik}}, \ and\ \bibinfo
  {author} {\bibfnamefont {T.}~\bibnamefont {Cagin}},\ }\href@noop {}
  {\bibfield  {journal} {\bibinfo  {journal} {Phys. Rev. B}\ }\textbf {\bibinfo
  {volume} {86}},\ \bibinfo {pages} {115410} (\bibinfo {year}
  {2012})}\BibitemShut {NoStop}%
\bibitem [{\citenamefont {Liu}\ \emph {et~al.}(2017)\citenamefont {Liu},
  \citenamefont {Ong}, \citenamefont {Wu}, \citenamefont {Zhao}, \citenamefont
  {Watanabe}, \citenamefont {Taniguchi}, \citenamefont {Chi}, \citenamefont
  {Zhang}, \citenamefont {Thong}, \citenamefont {Qiu} \emph
  {et~al.}}]{liu-BCN-2017}%
  \BibitemOpen
  \bibfield  {author} {\bibinfo {author} {\bibfnamefont {Y.}~\bibnamefont
  {Liu}}, \bibinfo {author} {\bibfnamefont {Z.-Y.}\ \bibnamefont {Ong}},
  \bibinfo {author} {\bibfnamefont {J.}~\bibnamefont {Wu}}, \bibinfo {author}
  {\bibfnamefont {Y.}~\bibnamefont {Zhao}}, \bibinfo {author} {\bibfnamefont
  {K.}~\bibnamefont {Watanabe}}, \bibinfo {author} {\bibfnamefont
  {T.}~\bibnamefont {Taniguchi}}, \bibinfo {author} {\bibfnamefont
  {D.}~\bibnamefont {Chi}}, \bibinfo {author} {\bibfnamefont {G.}~\bibnamefont
  {Zhang}}, \bibinfo {author} {\bibfnamefont {J.~T.}\ \bibnamefont {Thong}},
  \bibinfo {author} {\bibfnamefont {C.-W.}\ \bibnamefont {Qiu}},  \emph
  {et~al.},\ }\href@noop {} {\bibfield  {journal} {\bibinfo  {journal} {Sci.
  Rep.}\ }\textbf {\bibinfo {volume} {7}},\ \bibinfo {pages} {43886} (\bibinfo
  {year} {2017})}\BibitemShut {NoStop}%
\bibitem [{\citenamefont {Panchakarla}\ \emph {et~al.}(2009)\citenamefont
  {Panchakarla}, \citenamefont {Subrahmanyam}, \citenamefont {Saha},
  \citenamefont {Govindaraj}, \citenamefont {Krishnamurthy}, \citenamefont
  {Waghmare},\ and\ \citenamefont {Rao}}]{rao-BCN-2009}%
  \BibitemOpen
  \bibfield  {author} {\bibinfo {author} {\bibfnamefont {L.~S.}\ \bibnamefont
  {Panchakarla}}, \bibinfo {author} {\bibfnamefont {K.~S.}\ \bibnamefont
  {Subrahmanyam}}, \bibinfo {author} {\bibfnamefont {S.~K.}\ \bibnamefont
  {Saha}}, \bibinfo {author} {\bibfnamefont {A.}~\bibnamefont {Govindaraj}},
  \bibinfo {author} {\bibfnamefont {H.~R.}\ \bibnamefont {Krishnamurthy}},
  \bibinfo {author} {\bibfnamefont {U.~V.}\ \bibnamefont {Waghmare}}, \ and\
  \bibinfo {author} {\bibfnamefont {C.~N.~R.}\ \bibnamefont {Rao}},\
  }\href@noop {} {\bibfield  {journal} {\bibinfo  {journal} {Adv. Mater.}\
  }\textbf {\bibinfo {volume} {21}},\ \bibinfo {pages} {4726} (\bibinfo {year}
  {2009})}\BibitemShut {NoStop}%
\bibitem [{\citenamefont {Mann}\ \emph {et~al.}(2016)\citenamefont {Mann},
  \citenamefont {Rani}, \citenamefont {Kumar}, \citenamefont {Dubey},\ and\
  \citenamefont {Jindal}}]{jindal-2016}%
  \BibitemOpen
  \bibfield  {author} {\bibinfo {author} {\bibfnamefont {S.}~\bibnamefont
  {Mann}}, \bibinfo {author} {\bibfnamefont {P.}~\bibnamefont {Rani}}, \bibinfo
  {author} {\bibfnamefont {R.}~\bibnamefont {Kumar}}, \bibinfo {author}
  {\bibfnamefont {G.~S.}\ \bibnamefont {Dubey}}, \ and\ \bibinfo {author}
  {\bibfnamefont {V.}~\bibnamefont {Jindal}},\ }\href@noop {} {\bibfield
  {journal} {\bibinfo  {journal} {RSC Adv.}\ }\textbf {\bibinfo {volume} {6}},\
  \bibinfo {pages} {12158} (\bibinfo {year} {2016})}\BibitemShut {NoStop}%
\bibitem [{\citenamefont {Moses}\ \emph {et~al.}(2014)\citenamefont {Moses},
  \citenamefont {Shirodkar}, \citenamefont {Waghmare},\ and\ \citenamefont
  {Rao}}]{rao-BCN-2014}%
  \BibitemOpen
  \bibfield  {author} {\bibinfo {author} {\bibfnamefont {K.}~\bibnamefont
  {Moses}}, \bibinfo {author} {\bibfnamefont {S.~N.}\ \bibnamefont
  {Shirodkar}}, \bibinfo {author} {\bibfnamefont {U.~V.}\ \bibnamefont
  {Waghmare}}, \ and\ \bibinfo {author} {\bibfnamefont {C.~N.~R.}\ \bibnamefont
  {Rao}},\ }\href@noop {} {\bibfield  {journal} {\bibinfo  {journal} {Mater.
  Res. Express}\ }\textbf {\bibinfo {volume} {1}},\ \bibinfo {pages} {025603}
  (\bibinfo {year} {2014})}\BibitemShut {NoStop}%
\bibitem [{\citenamefont {Banerjee}\ and\ \citenamefont
  {Pati}(2014{\natexlab{b}})}]{pati-BCN-2014}%
  \BibitemOpen
  \bibfield  {author} {\bibinfo {author} {\bibfnamefont {S.}~\bibnamefont
  {Banerjee}}\ and\ \bibinfo {author} {\bibfnamefont {S.~K.}\ \bibnamefont
  {Pati}},\ }\href@noop {} {\bibfield  {journal} {\bibinfo  {journal}
  {Nanoscale}\ }\textbf {\bibinfo {volume} {6}},\ \bibinfo {pages} {13430}
  (\bibinfo {year} {2014}{\natexlab{b}})}\BibitemShut {NoStop}%
\bibitem [{\citenamefont {Shirodkar}\ and\ \citenamefont
  {Kaxiras}(2016)}]{shirodkar-BCN-2016}%
  \BibitemOpen
  \bibfield  {author} {\bibinfo {author} {\bibfnamefont {S.~N.}\ \bibnamefont
  {Shirodkar}}\ and\ \bibinfo {author} {\bibfnamefont {E.}~\bibnamefont
  {Kaxiras}},\ }\href@noop {} {\bibfield  {journal} {\bibinfo  {journal} {Phys.
  Rev. B}\ }\textbf {\bibinfo {volume} {93}},\ \bibinfo {pages} {245438}
  (\bibinfo {year} {2016})}\BibitemShut {NoStop}%
\bibitem [{\citenamefont {Banerjee}\ \emph {et~al.}(2016)\citenamefont
  {Banerjee}, \citenamefont {Neihsial},\ and\ \citenamefont
  {Pati}}]{pati-BCN-2016}%
  \BibitemOpen
  \bibfield  {author} {\bibinfo {author} {\bibfnamefont {S.}~\bibnamefont
  {Banerjee}}, \bibinfo {author} {\bibfnamefont {S.}~\bibnamefont {Neihsial}},
  \ and\ \bibinfo {author} {\bibfnamefont {S.~K.}\ \bibnamefont {Pati}},\
  }\href@noop {} {\bibfield  {journal} {\bibinfo  {journal} {J. Mater. Chem.
  A}\ }\textbf {\bibinfo {volume} {4}},\ \bibinfo {pages} {5517} (\bibinfo
  {year} {2016})}\BibitemShut {NoStop}%
\bibitem [{\citenamefont {Manna}\ and\ \citenamefont
  {Pati}(2011)}]{pati-BCN-2011}%
  \BibitemOpen
  \bibfield  {author} {\bibinfo {author} {\bibfnamefont {A.~K.}\ \bibnamefont
  {Manna}}\ and\ \bibinfo {author} {\bibfnamefont {S.~K.}\ \bibnamefont
  {Pati}},\ }\href@noop {} {\bibfield  {journal} {\bibinfo  {journal} {J. Phys.
  Chem. C}\ }\textbf {\bibinfo {volume} {115}},\ \bibinfo {pages} {10842}
  (\bibinfo {year} {2011})}\BibitemShut {NoStop}%
\bibitem [{\citenamefont {Singh}\ \emph {et~al.}(2013)\citenamefont {Singh},
  \citenamefont {Neek-Amal}, \citenamefont {Costamagna},\ and\ \citenamefont
  {Peeters}}]{peeters-BN-2013}%
  \BibitemOpen
  \bibfield  {author} {\bibinfo {author} {\bibfnamefont {S.~K.}\ \bibnamefont
  {Singh}}, \bibinfo {author} {\bibfnamefont {M.}~\bibnamefont {Neek-Amal}},
  \bibinfo {author} {\bibfnamefont {S.}~\bibnamefont {Costamagna}}, \ and\
  \bibinfo {author} {\bibfnamefont {F.~M.}\ \bibnamefont {Peeters}},\
  }\href@noop {} {\bibfield  {journal} {\bibinfo  {journal} {Phys. Rev. B}\
  }\textbf {\bibinfo {volume} {87}},\ \bibinfo {pages} {184106} (\bibinfo
  {year} {2013})}\BibitemShut {NoStop}%
\bibitem [{\citenamefont {Hong}\ \emph {et~al.}(2016)\citenamefont {Hong},
  \citenamefont {Zhang},\ and\ \citenamefont {Zeng}}]{hong-2016}%
  \BibitemOpen
  \bibfield  {author} {\bibinfo {author} {\bibfnamefont {Y.}~\bibnamefont
  {Hong}}, \bibinfo {author} {\bibfnamefont {J.}~\bibnamefont {Zhang}}, \ and\
  \bibinfo {author} {\bibfnamefont {X.~C.}\ \bibnamefont {Zeng}},\ }\href@noop
  {} {\bibfield  {journal} {\bibinfo  {journal} {Phys. Chem. Chem. Phys.}\
  }\textbf {\bibinfo {volume} {18}},\ \bibinfo {pages} {24164} (\bibinfo {year}
  {2016})}\BibitemShut {NoStop}%
\bibitem [{\citenamefont {Lindsay}\ and\ \citenamefont
  {Broido}(2010)}]{broido2010}%
  \BibitemOpen
  \bibfield  {author} {\bibinfo {author} {\bibfnamefont {L.}~\bibnamefont
  {Lindsay}}\ and\ \bibinfo {author} {\bibfnamefont {D.~A.}\ \bibnamefont
  {Broido}},\ }\href@noop {} {\bibfield  {journal} {\bibinfo  {journal} {Phys.
  Rev. B}\ }\textbf {\bibinfo {volume} {81}},\ \bibinfo {pages} {205441}
  (\bibinfo {year} {2010})}\BibitemShut {NoStop}%
\bibitem [{\citenamefont {Koukaras}\ \emph {et~al.}(2015)\citenamefont
  {Koukaras}, \citenamefont {Kalosakas}, \citenamefont {Galiotis},\ and\
  \citenamefont {Papagelis}}]{koukaras-srep-2015}%
  \BibitemOpen
  \bibfield  {author} {\bibinfo {author} {\bibfnamefont {E.~N.}\ \bibnamefont
  {Koukaras}}, \bibinfo {author} {\bibfnamefont {G.}~\bibnamefont {Kalosakas}},
  \bibinfo {author} {\bibfnamefont {C.}~\bibnamefont {Galiotis}}, \ and\
  \bibinfo {author} {\bibfnamefont {K.}~\bibnamefont {Papagelis}},\ }\href@noop
  {} {\bibfield  {journal} {\bibinfo  {journal} {Sci. Rep.}\ }\textbf {\bibinfo
  {volume} {5}},\ \bibinfo {pages} {12923} (\bibinfo {year}
  {2015})}\BibitemShut {NoStop}%
\bibitem [{\citenamefont {Plimpton}(1995)}]{lammps}%
  \BibitemOpen
  \bibfield  {author} {\bibinfo {author} {\bibfnamefont {S.}~\bibnamefont
  {Plimpton}},\ }\href@noop {} {\bibfield  {journal} {\bibinfo  {journal} {J.
  Comput. Phys.}\ }\textbf {\bibinfo {volume} {117}},\ \bibinfo {pages} {1}
  (\bibinfo {year} {1995})}\BibitemShut {NoStop}%
\bibitem [{\citenamefont {Martyna}\ \emph {et~al.}(1994)\citenamefont
  {Martyna}, \citenamefont {Tobias},\ and\ \citenamefont
  {Klein}}]{martyna-1994}%
  \BibitemOpen
  \bibfield  {author} {\bibinfo {author} {\bibfnamefont {G.~J.}\ \bibnamefont
  {Martyna}}, \bibinfo {author} {\bibfnamefont {D.~J.}\ \bibnamefont {Tobias}},
  \ and\ \bibinfo {author} {\bibfnamefont {M.~L.}\ \bibnamefont {Klein}},\
  }\href@noop {} {\bibfield  {journal} {\bibinfo  {journal} {J. Chem. Phys.}\
  }\textbf {\bibinfo {volume} {101}},\ \bibinfo {pages} {4177} (\bibinfo {year}
  {1994})}\BibitemShut {NoStop}%
\bibitem [{\citenamefont {Ci}\ \emph {et~al.}(2010)\citenamefont {Ci},
  \citenamefont {Song}, \citenamefont {Jin}, \citenamefont {Jariwala},
  \citenamefont {Wu}, \citenamefont {Li}, \citenamefont {Srivastava},
  \citenamefont {Wang}, \citenamefont {Storr}, \citenamefont {Balicas},
  \citenamefont {Liu},\ and\ \citenamefont {Ajayan}}]{ajayan-2011-nmat}%
  \BibitemOpen
  \bibfield  {author} {\bibinfo {author} {\bibfnamefont {L.}~\bibnamefont
  {Ci}}, \bibinfo {author} {\bibfnamefont {L.}~\bibnamefont {Song}}, \bibinfo
  {author} {\bibfnamefont {C.}~\bibnamefont {Jin}}, \bibinfo {author}
  {\bibfnamefont {D.}~\bibnamefont {Jariwala}}, \bibinfo {author}
  {\bibfnamefont {D.}~\bibnamefont {Wu}}, \bibinfo {author} {\bibfnamefont
  {Y.}~\bibnamefont {Li}}, \bibinfo {author} {\bibfnamefont {A.}~\bibnamefont
  {Srivastava}}, \bibinfo {author} {\bibfnamefont {Z.}~\bibnamefont {Wang}},
  \bibinfo {author} {\bibfnamefont {K.}~\bibnamefont {Storr}}, \bibinfo
  {author} {\bibfnamefont {L.}~\bibnamefont {Balicas}}, \bibinfo {author}
  {\bibfnamefont {F.}~\bibnamefont {Liu}}, \ and\ \bibinfo {author}
  {\bibfnamefont {P.~M.}\ \bibnamefont {Ajayan}},\ }\href@noop {} {\bibfield
  {journal} {\bibinfo  {journal} {Nat. Mater.}\ }\textbf {\bibinfo {volume}
  {9}},\ \bibinfo {pages} {430} (\bibinfo {year} {2010})}\BibitemShut {NoStop}%
\bibitem [{\citenamefont {Anees}\ \emph {et~al.}(2016)\citenamefont {Anees},
  \citenamefont {Valsakumar},\ and\ \citenamefont
  {Panigrahia}}]{valsakumar2016}%
  \BibitemOpen
  \bibfield  {author} {\bibinfo {author} {\bibfnamefont {P.}~\bibnamefont
  {Anees}}, \bibinfo {author} {\bibfnamefont {M.~C.}\ \bibnamefont
  {Valsakumar}}, \ and\ \bibinfo {author} {\bibfnamefont {B.~K.}\ \bibnamefont
  {Panigrahia}},\ }\href@noop {} {\bibfield  {journal} {\bibinfo  {journal}
  {Phys. Chem. Chem. Phys.}\ }\textbf {\bibinfo {volume} {18}},\ \bibinfo
  {pages} {2672} (\bibinfo {year} {2016})}\BibitemShut {NoStop}%
\bibitem [{\citenamefont {Seol}\ \emph {et~al.}(2010)\citenamefont {Seol},
  \citenamefont {Jo}, \citenamefont {Moore}, \citenamefont {Lindsay},
  \citenamefont {Aitken}, \citenamefont {Pettes}, \citenamefont {Li},
  \citenamefont {Yao}, \citenamefont {Huang}, \citenamefont {Broido},
  \citenamefont {Mingo}, \citenamefont {Ruoff},\ and\ \citenamefont
  {Shi}}]{seol-science-2010}%
  \BibitemOpen
  \bibfield  {author} {\bibinfo {author} {\bibfnamefont {J.~H.}\ \bibnamefont
  {Seol}}, \bibinfo {author} {\bibfnamefont {I.}~\bibnamefont {Jo}}, \bibinfo
  {author} {\bibfnamefont {A.~L.}\ \bibnamefont {Moore}}, \bibinfo {author}
  {\bibfnamefont {L.}~\bibnamefont {Lindsay}}, \bibinfo {author} {\bibfnamefont
  {Z.~H.}\ \bibnamefont {Aitken}}, \bibinfo {author} {\bibfnamefont {M.~T.}\
  \bibnamefont {Pettes}}, \bibinfo {author} {\bibfnamefont {X.}~\bibnamefont
  {Li}}, \bibinfo {author} {\bibfnamefont {Z.}~\bibnamefont {Yao}}, \bibinfo
  {author} {\bibfnamefont {R.}~\bibnamefont {Huang}}, \bibinfo {author}
  {\bibfnamefont {D.}~\bibnamefont {Broido}}, \bibinfo {author} {\bibfnamefont
  {N.}~\bibnamefont {Mingo}}, \bibinfo {author} {\bibfnamefont {R.~S.}\
  \bibnamefont {Ruoff}}, \ and\ \bibinfo {author} {\bibfnamefont
  {L.}~\bibnamefont {Shi}},\ }\href@noop {} {\bibfield  {journal} {\bibinfo
  {journal} {Science}\ }\textbf {\bibinfo {volume} {328}},\ \bibinfo {pages}
  {213} (\bibinfo {year} {2010})}\BibitemShut {NoStop}%
\bibitem [{\citenamefont {Lindsay}\ \emph {et~al.}(2010)\citenamefont
  {Lindsay}, \citenamefont {Broido},\ and\ \citenamefont
  {Mingo}}]{lindsay-PRB-2010}%
  \BibitemOpen
  \bibfield  {author} {\bibinfo {author} {\bibfnamefont {L.}~\bibnamefont
  {Lindsay}}, \bibinfo {author} {\bibfnamefont {D.}~\bibnamefont {Broido}}, \
  and\ \bibinfo {author} {\bibfnamefont {N.}~\bibnamefont {Mingo}},\
  }\href@noop {} {\bibfield  {journal} {\bibinfo  {journal} {Phys. Rev. B}\
  }\textbf {\bibinfo {volume} {82}},\ \bibinfo {pages} {115427} (\bibinfo
  {year} {2010})}\BibitemShut {NoStop}%
\bibitem [{\citenamefont {Calizo}\ \emph {et~al.}(2007)\citenamefont {Calizo},
  \citenamefont {Balandin}, \citenamefont {Bao}, \citenamefont {Miao},\ and\
  \citenamefont {Lau}}]{balandin-nl-2007}%
  \BibitemOpen
  \bibfield  {author} {\bibinfo {author} {\bibfnamefont {I.}~\bibnamefont
  {Calizo}}, \bibinfo {author} {\bibfnamefont {A.}~\bibnamefont {Balandin}},
  \bibinfo {author} {\bibfnamefont {W.}~\bibnamefont {Bao}}, \bibinfo {author}
  {\bibfnamefont {F.}~\bibnamefont {Miao}}, \ and\ \bibinfo {author}
  {\bibfnamefont {C.}~\bibnamefont {Lau}},\ }\href@noop {} {\bibfield
  {journal} {\bibinfo  {journal} {Nano Lett.}\ }\textbf {\bibinfo {volume}
  {7}},\ \bibinfo {pages} {2645} (\bibinfo {year} {2007})}\BibitemShut
  {NoStop}%
\bibitem [{\citenamefont {Zwanzig}(1965)}]{zwanzig-1965}%
  \BibitemOpen
  \bibfield  {author} {\bibinfo {author} {\bibfnamefont {R.}~\bibnamefont
  {Zwanzig}},\ }\href@noop {} {\bibfield  {journal} {\bibinfo  {journal} {Annu.
  Rev. Phys. Chem.}\ }\textbf {\bibinfo {volume} {16}},\ \bibinfo {pages} {67}
  (\bibinfo {year} {1965})}\BibitemShut {NoStop}%
\bibitem [{\citenamefont {McQuarrie}(2000)}]{McQuarrie_statmech_book}%
  \BibitemOpen
  \bibfield  {author} {\bibinfo {author} {\bibfnamefont {D.~A.}\ \bibnamefont
  {McQuarrie}},\ }\href {https://books.google.com/books?id=itcpPnDnJM0C} {\emph
  {\bibinfo {title} {Statistical Mechanics}}}\ (\bibinfo  {publisher}
  {University Science Books},\ \bibinfo {year} {2000})\BibitemShut {NoStop}%
\bibitem [{\citenamefont {Hong}\ \emph {et~al.}(2015)\citenamefont {Hong},
  \citenamefont {Zhang}, \citenamefont {Huang},\ and\ \citenamefont
  {Zeng}}]{xczeng2015}%
  \BibitemOpen
  \bibfield  {author} {\bibinfo {author} {\bibfnamefont {Y.}~\bibnamefont
  {Hong}}, \bibinfo {author} {\bibfnamefont {J.}~\bibnamefont {Zhang}},
  \bibinfo {author} {\bibfnamefont {X.}~\bibnamefont {Huang}}, \ and\ \bibinfo
  {author} {\bibfnamefont {X.~C.}\ \bibnamefont {Zeng}},\ }\href@noop {}
  {\bibfield  {journal} {\bibinfo  {journal} {Nanoscale}\ }\textbf {\bibinfo
  {volume} {7}},\ \bibinfo {pages} {18716} (\bibinfo {year}
  {2015})}\BibitemShut {NoStop}%
\bibitem [{\citenamefont {Lee}\ \emph {et~al.}(1991)\citenamefont {Lee},
  \citenamefont {Biswas}, \citenamefont {Soukoulis}, \citenamefont {Wang},
  \citenamefont {Chan},\ and\ \citenamefont {Ho}}]{lee-1991}%
  \BibitemOpen
  \bibfield  {author} {\bibinfo {author} {\bibfnamefont {Y.~H.}\ \bibnamefont
  {Lee}}, \bibinfo {author} {\bibfnamefont {R.}~\bibnamefont {Biswas}},
  \bibinfo {author} {\bibfnamefont {C.~M.}\ \bibnamefont {Soukoulis}}, \bibinfo
  {author} {\bibfnamefont {C.~Z.}\ \bibnamefont {Wang}}, \bibinfo {author}
  {\bibfnamefont {C.~T.}\ \bibnamefont {Chan}}, \ and\ \bibinfo {author}
  {\bibfnamefont {K.~M.}\ \bibnamefont {Ho}},\ }\href@noop {} {\bibfield
  {journal} {\bibinfo  {journal} {Phys. Rev. B}\ }\textbf {\bibinfo {volume}
  {43}},\ \bibinfo {pages} {6573} (\bibinfo {year} {1991})}\BibitemShut
  {NoStop}%
\bibitem [{\citenamefont {Haskins}\ \emph {et~al.}(2014)\citenamefont
  {Haskins}, \citenamefont {Kinaci}, \citenamefont {Sevik},\ and\ \citenamefont
  {Cagin}}]{tahirjcp2014}%
  \BibitemOpen
  \bibfield  {author} {\bibinfo {author} {\bibfnamefont {J.~B.}\ \bibnamefont
  {Haskins}}, \bibinfo {author} {\bibfnamefont {A.}~\bibnamefont {Kinaci}},
  \bibinfo {author} {\bibfnamefont {C.}~\bibnamefont {Sevik}}, \ and\ \bibinfo
  {author} {\bibfnamefont {T.}~\bibnamefont {Cagin}},\ }\href@noop {}
  {\bibfield  {journal} {\bibinfo  {journal} {J. Chem. Phys.}\ }\textbf
  {\bibinfo {volume} {140}},\ \bibinfo {pages} {244112} (\bibinfo {year}
  {2014})}\BibitemShut {NoStop}%
\bibitem [{\citenamefont {Khan}\ \emph {et~al.}(2015)\citenamefont {Khan},
  \citenamefont {Navid}, \citenamefont {Noshin}, \citenamefont {Uddin},
  \citenamefont {Hossain},\ and\ \citenamefont
  {Subrina}}]{size-dependence-tersoff-2015}%
  \BibitemOpen
  \bibfield  {author} {\bibinfo {author} {\bibfnamefont {A.~I.}\ \bibnamefont
  {Khan}}, \bibinfo {author} {\bibfnamefont {I.~A.}\ \bibnamefont {Navid}},
  \bibinfo {author} {\bibfnamefont {M.}~\bibnamefont {Noshin}}, \bibinfo
  {author} {\bibfnamefont {H.~M.~A.}\ \bibnamefont {Uddin}}, \bibinfo {author}
  {\bibfnamefont {F.~F.}\ \bibnamefont {Hossain}}, \ and\ \bibinfo {author}
  {\bibfnamefont {S.}~\bibnamefont {Subrina}},\ }\href@noop {} {\bibfield
  {journal} {\bibinfo  {journal} {Electronics}\ }\textbf {\bibinfo {volume}
  {4}},\ \bibinfo {pages} {1109} (\bibinfo {year} {2015})}\BibitemShut
  {NoStop}%
\bibitem [{\citenamefont {Khan}\ \emph {et~al.}(2017)\citenamefont {Khan},
  \citenamefont {Navid}, \citenamefont {Noshin},\ and\ \citenamefont
  {Subrina}}]{KhanBN2017}%
  \BibitemOpen
  \bibfield  {author} {\bibinfo {author} {\bibfnamefont {A.~I.}\ \bibnamefont
  {Khan}}, \bibinfo {author} {\bibfnamefont {I.~A.}\ \bibnamefont {Navid}},
  \bibinfo {author} {\bibfnamefont {M.}~\bibnamefont {Noshin}}, \ and\ \bibinfo
  {author} {\bibfnamefont {S.}~\bibnamefont {Subrina}},\ }\href {\doibase
  10.1063/1.4997036} {\bibfield  {journal} {\bibinfo  {journal} {AIP Advances}\
  }\textbf {\bibinfo {volume} {7}},\ \bibinfo {pages} {105110} (\bibinfo {year}
  {2017})}\BibitemShut {NoStop}%
\bibitem [{\citenamefont {Wang}\ \emph {et~al.}(1990)\citenamefont {Wang},
  \citenamefont {Chan},\ and\ \citenamefont {Ho}}]{wang-PRB-1990}%
  \BibitemOpen
  \bibfield  {author} {\bibinfo {author} {\bibfnamefont {C.~Z.}\ \bibnamefont
  {Wang}}, \bibinfo {author} {\bibfnamefont {C.~T.}\ \bibnamefont {Chan}}, \
  and\ \bibinfo {author} {\bibfnamefont {K.~M.}\ \bibnamefont {Ho}},\
  }\href@noop {} {\bibfield  {journal} {\bibinfo  {journal} {Phys. Rev. B}\
  }\textbf {\bibinfo {volume} {42}},\ \bibinfo {pages} {11276} (\bibinfo {year}
  {1990})}\BibitemShut {NoStop}%
\bibitem [{\citenamefont {Che}\ \emph {et~al.}(2000)\citenamefont {Che},
  \citenamefont {Cagin}, \citenamefont {Deng},\ and\ \citenamefont
  {Goddard~III}}]{tahir-JCP-2000}%
  \BibitemOpen
  \bibfield  {author} {\bibinfo {author} {\bibfnamefont {J.}~\bibnamefont
  {Che}}, \bibinfo {author} {\bibfnamefont {T.}~\bibnamefont {Cagin}}, \bibinfo
  {author} {\bibfnamefont {W.}~\bibnamefont {Deng}}, \ and\ \bibinfo {author}
  {\bibfnamefont {W.~A.}\ \bibnamefont {Goddard~III}},\ }\href@noop {}
  {\bibfield  {journal} {\bibinfo  {journal} {J. Chem. Phys.}\ }\textbf
  {\bibinfo {volume} {113}},\ \bibinfo {pages} {6888} (\bibinfo {year}
  {2000})}\BibitemShut {NoStop}%
\end{thebibliography}%
\end{document}